\documentclass[letterpaper,twocolumn,ngerman,english,aps, pra, showpacs]{revtex4}
\pdfoutput=1
\usepackage[T1]{fontenc}
\usepackage[utf8]{inputenc}
\usepackage{subfigure}%
\usepackage{verbatim}
\usepackage{amsmath}%
\usepackage{color}%
\usepackage{graphicx}%
\usepackage{amssymb}%
\usepackage{esint}%

\makeatletter

\usepackage{babel}
\makeatother

\begin{document}

\newcommand{\dd}{\mathrm{d}}

\newcommand{\bra}[1]{\left<#1\right|}

\newcommand{\ket}[1]{\left|#1\right>}

\newcommand{\braket}[2]{\left<#1\vphantom{#2}\right|\left.\hspace{-0.25em}#2\vphantom{#1}\right>}

\title{Adaptive estimation of qubits \\
by symmetry measurements}

\author{Christof J. Happ and Matthias Freyberger}

\affiliation{Institut f\"ur Quantenphysik, Universit\"at Ulm, D-89069 Ulm, Germany}

\begin{abstract}
We analyze quantum state estimation for finite samples based on symmetry information.
The used measurement concept compares an unknown qubit to a reference
state. We describe explicitly an adaptive strategy, that enhances
the estimation fidelity of these measurements.
\end{abstract}


\maketitle

\section{Introduction}

The concept of a measurement lies at the heart of quantum mechanics.
For a long time this has been a topic of fundamental discussion \citep{WheelerZureck}.
Quantum information science \citep{QIV} has transformed it to a basic
technical issue, at the core of many applications \citep{BennetEtAl93}.
Of particular importance are quantum state measurements, since they
can reveal complete information about a quantum system, or more precisely
about its preparation process \citep{Fano57,Peres84}. In this sense
such measurement schemes attach a certain operational meaning to a
quantum state. 

Full tomography \citep{SR97} of an unknown state requires infinitely
many identical copies of a quantum system. The no-cloning theorem
\citep{NoClone}, however, prevents to produce them from a single
system. Hence we always encounter the typical situation of an estimation
approach \citep{vandenBos,ParisRehacek}: It is only possible to collect
a \emph{limited} amount of information from a \emph{finite} sample
of identical systems. Using the corresponding finite set of measurement
data, we then have to deliver an estimate for the underlying quantum
state. The benchmark for the quality of such a procedure is the optimal
average fidelity of a joint measurement \citep{BM99,MP95,DBE98} on
all systems of the finite sample. On the other hand, the needed physical
principles for such collective measurements are not simple to realize.

Hence this optimal approach has been approximated by various local
measurement schemes \citep{FKF00,BM02-04,GM00,Jones94,HannemannEtAl02,EN04}
combined with suitable classical communication. The physics of the
related measurements on single systems has a rather simple operational
meaning, while the obtained estimation fidelities can be close to
the optimal ones.

The present contribution aims at the estimation of a pure qubit of
which only a limited number $N$ of identically prepared copies is
available. Our aim is not to optimize a certain fidelity. We rather
examine how a basic measurement concept can be used for estimation
of quantum states. We investigate measurements that compare the unknown
copies, one by one, to a known reference state, a {}``quantum ruler''.
For these comparisons we restrict ourselves to measurements that yield
symmetry information only. Such measurements have been used in \citep{BCJ03}
for state comparison, and based on this work applied to programmable
unambiguous state discrimination \citep{BH05}. Moreover these symmetry
measurements are a simple realization of a quantum {}``multimeter''
\citep{FDF02}, where the ruler state is {}``quantum software''
that defines what measurement is performed.

\section{Basic concepts}

\subsection{Symmetry measurements}

To be more specific, our comparison is done by symmetry measurements.
In contrast to a spin measurement such a comparison is close to classical
concepts of measurement: We determine the agreement between the unknown
system and a known ruler. Therefore, it will be interesting to see
how successful this ansatz can be in the quantum domain.

In order to define this in more detail, we write the unknown qubit
state\begin{equation}
\left|\psi\right\rangle =\cos\frac{\theta}{2}\left|0\right\rangle +e^{i\phi}\sin\frac{\theta}{2}\left|1\right\rangle \label{eq:psi}\end{equation}
in a computational basis $\left\{ \ket{0},\ket{1}\right\} $. It is
assumed that $N$ identically prepared copies of this qubit are available,
i.\,e. they all stem from the same but unknown preparation process.
Each copy is compared to a reference state\begin{equation}
\left|r_{\nu}\right\rangle =\cos\frac{\vartheta_{\nu}}{2}\left|0\right\rangle +e^{i\varphi_{\nu}}\sin\frac{\vartheta_{\nu}}{2}\left|1\right\rangle ,\label{eq:rnu}\end{equation}
with $\nu=1,...,N$. Hence for each measurement step $\nu$ the corresponding
two-particle system is in the product state $\left|\psi\right\rangle \left|r_{\nu}\right\rangle $.
Our measurement concept now consists in determining the symmetry of
this state with respect to particle exchange. The results will be
dichotomic: either we obtain a {}``symmetric ($s$)'' or an {}``anti\-symmetric
($a$)'' signature. 

The probability of finding the product state in the antisymmetric subspace,
spanned by the antisymmetric state $\ket{\Psi^{-}}=\frac{1}{\sqrt{2}}\left(\left|01\right\rangle -\left|10\right\rangle \right)$,
reads\begin{align}
 & p_{a}(\left|\psi\right\rangle ,\left|r_{\nu}\right\rangle )=\nonumber \\
 & =\frac{1}{4}\left(1-\cos\theta\cos\vartheta_{\nu}-\cos\left(\phi-\varphi_{\nu}\right)\sin\theta\sin\vartheta_{\nu}\right).\label{eq:pa}\end{align}
We denote the complementary probability, i.\,e. the probability of
finding the product state in the symmetric subspace, by $p_{s}(\left|\psi\right\rangle ,\left|r_{\nu}\right\rangle )=1-p_{a}(\left|\psi\right\rangle ,\left|r_{\nu}\right\rangle )$.

Before discussing how to utilize this kind of measurement for quantum
state estimation, we mention a simple realization by linear optics
\citep{BellTele}. Let us assume, that the unknown and the reference
qubit are prepared as superpositions of single photon polarizations.
These photons impinge on a 50:50 beam splitter that is perfectly balanced
concerning the two polarizations in each of its input ports. That
is, the unknown photon enters one input port, the reference photon
the other one. The probability of finding a single photon with arbitrary
polarization in each output port is then equivalent to $p_{a}$, Eq.\,(\ref{eq:pa}).
We therefore can measure it with efficient detectors that simply distinguish
the vacuum from any finite number of photons. This measurement scheme
has already been experimentally utilized for measuring the overlap
of quantum states \citep{HDFF03}.

\subsection{Estimation method}

We perform symmetry measurements on $N$ copies of the qubit $\left|\psi\right\rangle $
and the corresponding reference states. For each single reference
$\left|r_{\nu}\right\rangle $ we obtain a measurement result $\alpha_{\nu}$,
which is either $a$, for the antisymmetric subspace, or $s$, for
the symmetric subspace, respectively. According to Bayes' rule \citep{Harney,SBC01}
we can calculate the probability of this sequence $\mathbb{S}_{N}=\{\alpha_{\nu},\ket{r_{\nu}};\nu\in[1,N]\}$
of results and references, and use it as a likelihood function \citep{ParisRehacek,Hradil97}
\begin{equation}
L_{N}=\prod_{\nu=1}^{N}p_{\alpha_{\nu}}(\ket{\psi},\left|r_{\nu}\right\rangle )\label{eq:Like}\end{equation}
for the unknown qubit $\left|\psi\right\rangle $. Note that $L_{N}$
is given as a function of the parameters $\theta$ and $\phi$, according
to Eq.\,(\ref{eq:psi}). Their values at the maximum of $L_{N}$
determine the estimated state $\ket{\psi_{N}^{\mathrm{est}}}$ for
one specific measurement sequence $\mathbb{S}_{N}$ \citep{FootMixed}.

So far we still have no rule how to choose the reference states. The
first reference $\left|r_{1}\right\rangle $ is always arbitrary,
since we have no information about the state $\ket{\psi}$ before
the first measurement. We therefore always choose $\left|r_{1}\right\rangle =\ket{0}$.
But at a later measurement stage we already have the results of the
preceding measurements. Then an adaptive strategy for choosing the
new references using the already obtained data is possible and will
be described in the next section. After fixing the next reference
according to a chosen adaption strategy, we perform the next symmetry
measurement. Then we repeat this cycle until all copies of the unknown
state are used up. 

To quantify how well the unknown state $\ket{\psi}$ and its final
estimate $\left|\psi_{N}^{\mathrm{est}}\right\rangle $ coincide,
we apply the corresponding fidelity $F_{N}=\left|\braket{\psi_{N}^{\mathrm{est}}}{\psi}\right|^{2}$.
This value of course depends on $\ket{\psi}$ as well as on the probabilistic
outcomes $\left\{ \alpha_{\nu}\right\} $ in the measurement sequence
$\mathbb{S}_{N}$. To arrive at a measure for the quality of the used
adaptive measurement strategy we have to average over all unknown
states $\ket{\psi}$ on the Bloch sphere, and all measurement outcomes.
This leads to the mean expected estimation fidelity \begin{equation}
\left\langle F_{N}\right\rangle =\int\left\langle \left|\braket{\psi_{N}^{\mathrm{est}}}{\psi}\right|^{2}\right\rangle _{\mathbb{S}_{N}}\mathrm{d}\Omega_{\psi},\label{eq:Fquerex}\end{equation}
where the integration $\int\mathrm{d}\Omega_{\psi}=\frac{1}{4\pi}\int_{0}^{2\pi}\dd\phi\int_{0}^{\pi}\dd\theta\sin\theta$
denotes a Bloch sphere average and $\left\langle ...\right\rangle _{\mathbb{S}_{N}}$
is the average over all measurement sequences possible within the
chosen adaption strategy.

\section{Adaption strategy}

\begin{figure}
\includegraphics[clip,angle=-90,width=1\columnwidth]{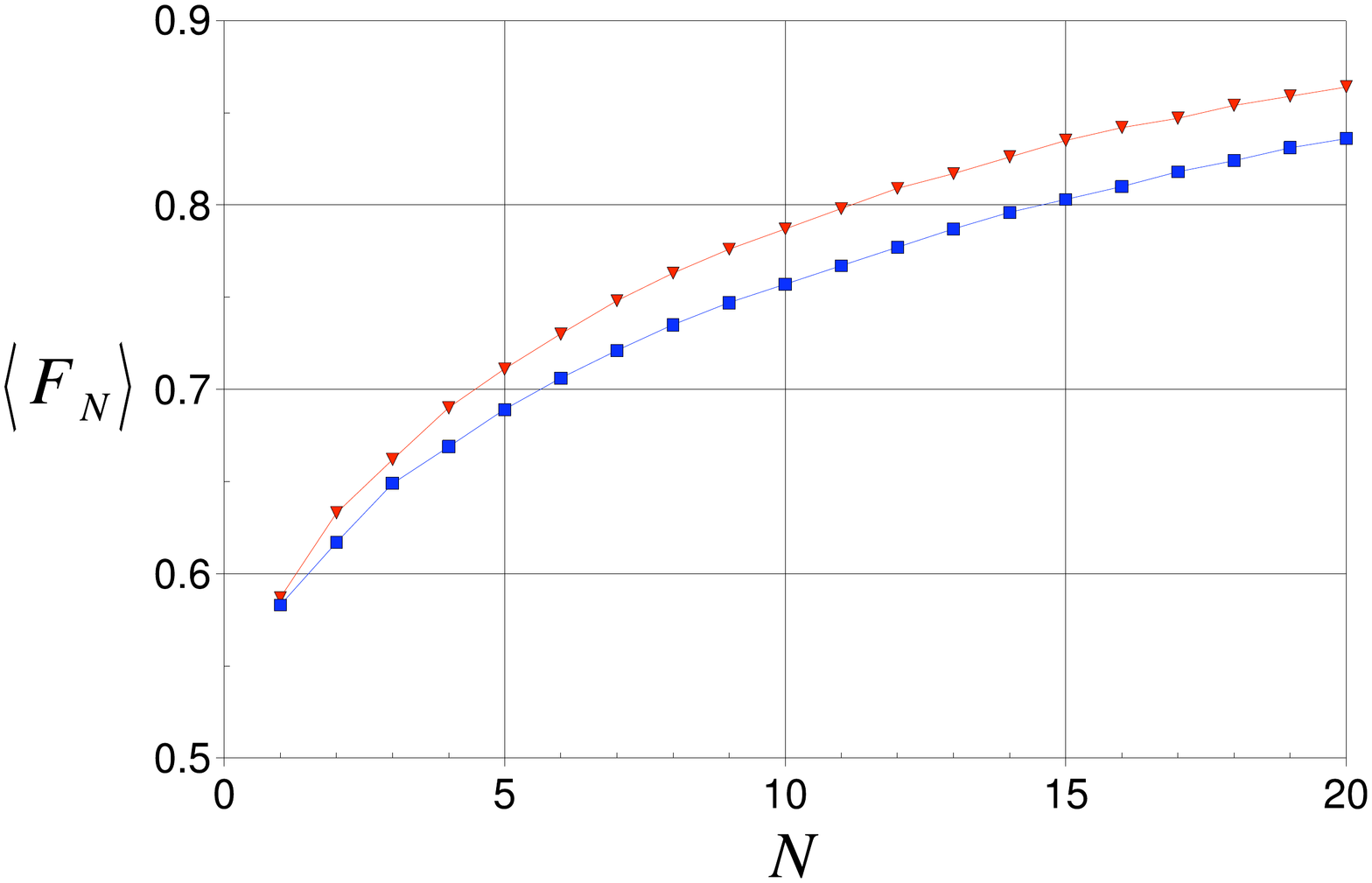}

\caption{ \emph{Fidelities from symmetry measurements}\label{fig:Fidelity-maximization}.
The achieved mean estimation fidelity $\left\langle F_{N}\right\rangle $,
Eq.\,(\ref{eq:Fquerex}), is plotted over the number $N$ of performed
measurements, i.\,e. the number of copies of the unknown qubit. Using
an adaptive method (symbols~\textcolor{red}{$\blacktriangledown$})
gives fidelities which are up to $5\%$ above those obtained with
random references (symbols~\textcolor{blue}{$\blacksquare$}). 
Note that these numerically simulated data points have a small error 
due to the finite amount of Monte Carlo runs.
However, this error is smaller than the size of the symbols, 
as can be seen for the $N=1$ data points, which should in principle coincide}

\end{figure}
\begin{figure}
\subfigure[$N=1$]{\includegraphics[bb=154 595 552 719, width=1\columnwidth]{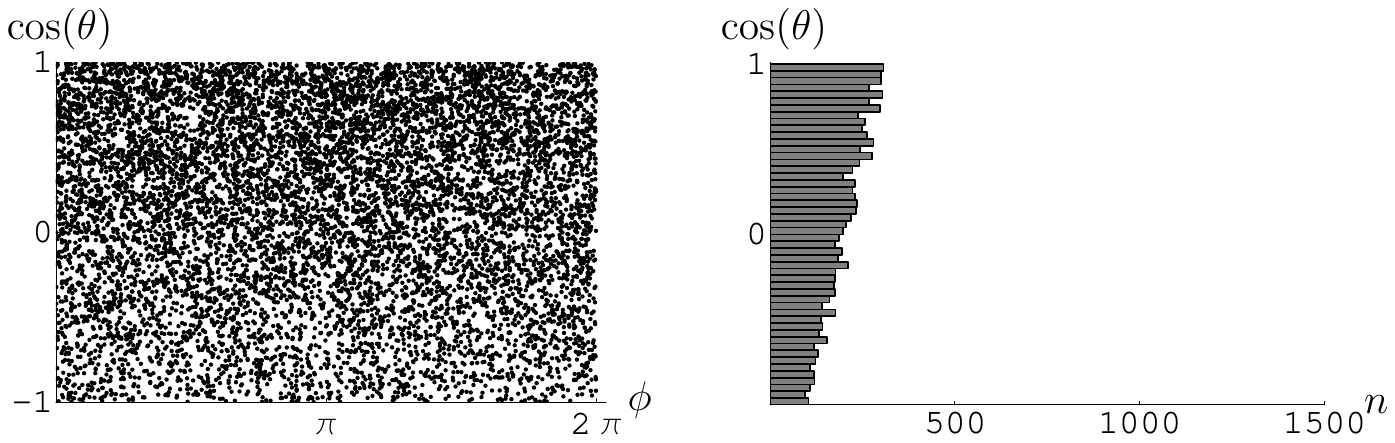}}

\subfigure[$N=5$]{\includegraphics[bb=154 595 552 719, width=1\columnwidth]{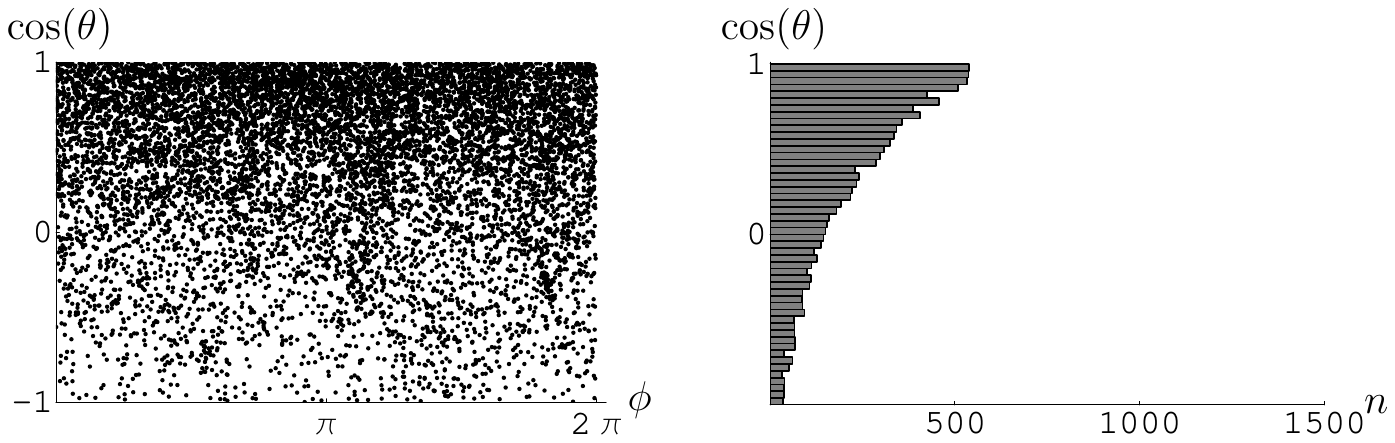}}

\subfigure[$N=20$]{\includegraphics[bb=154 595 552 719, width=1\columnwidth]{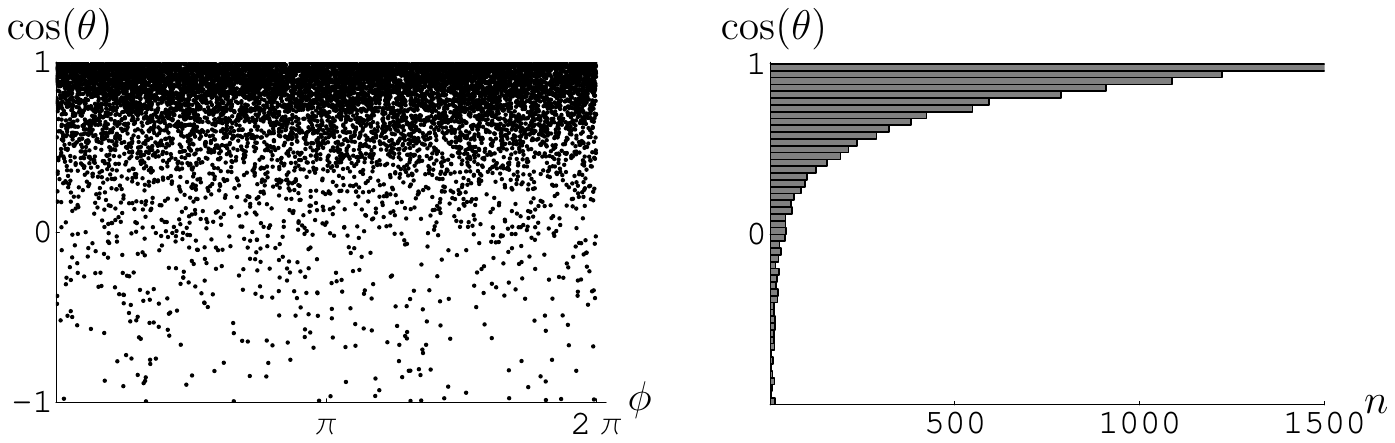}}

\caption{\emph{Estimated states on the Bloch sphere.\label{fig:EstBK}} On
the left hand side we depict the Bloch sphere distribution of estimated
states for $10^{4}$ different runs and three sizes $N$ of the finite
ensemble. Each point $(\theta_{N}^{\mathrm{est}},\phi_{N}^{\mathrm{est}})$
corresponds to one simulated measurement run, where the prepared,
unknown state has been chosen randomly. In order to see the convergence,
the coordinate frame has been rotated after the estimation such that
the unknown state would lie on the north pole, i.\,e. $\cos\theta_{}^{\mathrm{}}=1$.
On the right hand side the same data sets are shown as histograms
for $\cos\theta_{}^{}$. The convergence of the estimated states towards
the correct state can be clearly seen.}

\end{figure}
\begin{figure}[t]
\subfigure[$N=1$]{\includegraphics[bb=154 595 552 719, width=1\columnwidth]{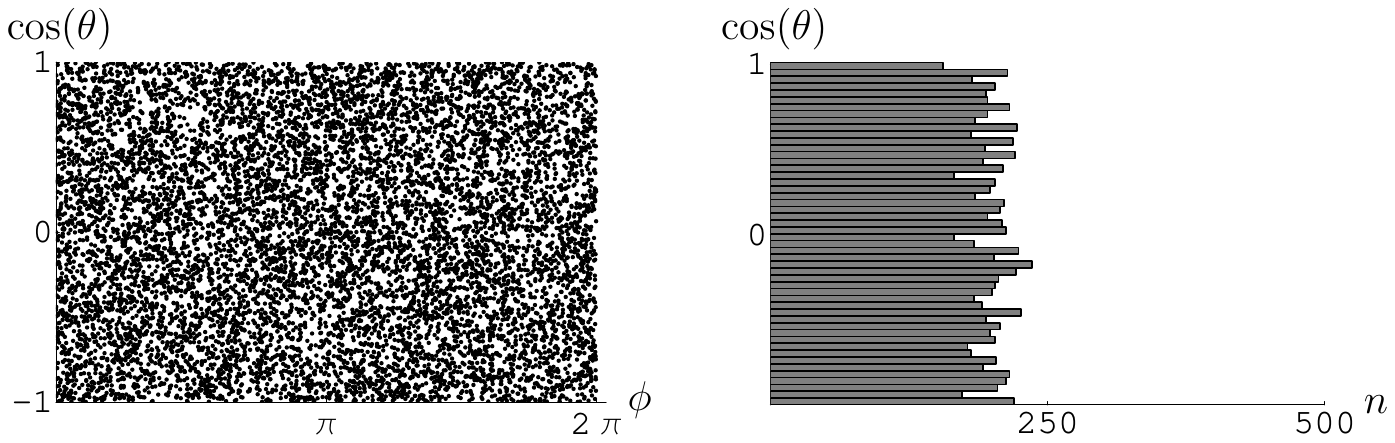}}

\subfigure[$N=5$]{\includegraphics[bb=154 595 552 719, width=1\columnwidth]{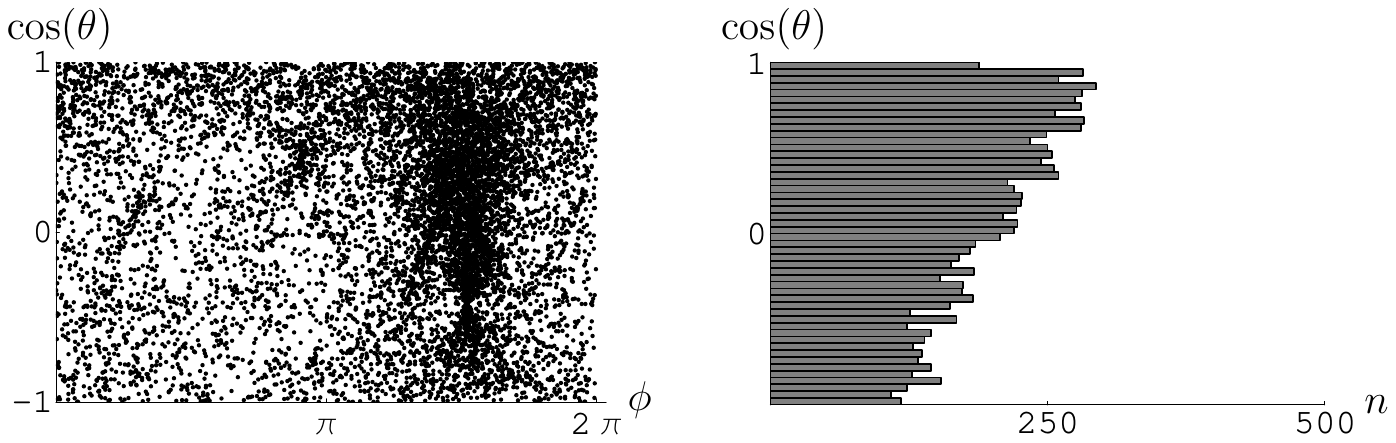}}

\subfigure[$N=20$]{\includegraphics[bb=154 595 552 719, width=1\columnwidth]{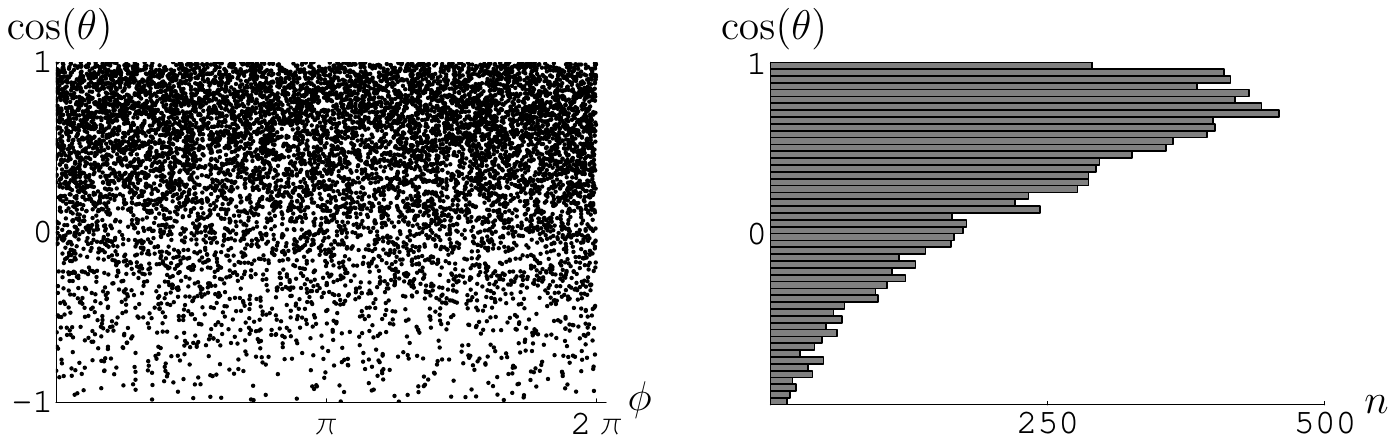}}

\caption{\emph{Reference states on the Bloch sphere.\label{fig:RefBK}} Here
we show distributions of reference states, Eq.\,(\ref{eq:rnu}),
over the Bloch sphere and corresponding histograms, again rotated
as described in the caption of Fig.\,\ref{fig:EstBK}. 
Due to this rotation the first reference states $\ket{r_{1}}=\ket{0}$ (in the calculational basis)
are randomly distributed (a) with respect to the unknown state.
It can be seen, that for higher numbers of measurements
the references tend to be close to the prepared state, but they do
not become completely parallel. Also, in comparison to Fig.\,\ref{fig:EstBK}
the convergence is weaker. The azimuthal asymmetry in (b) is due to
the small number of possible adaptions for the reference state $\ket{r_{5}}$.
After four steps we can obtain one of $2^{4}$ possible sets of measurement
data. Therefore, there are only $16$ possibilities for the orientation
of the reference $\ket{r_{5}}$ relative to $\ket{r_{1}}$. As can
be seen, this breaks the equipartition over the $\varphi_{5}$ coordinate. }

\end{figure}
The simplest way to construct an estimation scheme, which achieves
equal average fidelities for all possible unknown states, is to use
equally distributed random reference states. Of course this is not
an adaption, but a lower bound to which adaptive schemes can be compared
in order to assess their quality. To calculate the corresponding fidelity,
Eq. (\ref{eq:Fquerex}), we use Monte Carlo techniques. That means
the result of a measurement is simulated by a random number generator, 
in such a way that the results follow the distribution Eq.\,(\ref{eq:pa}).
From these randomized results we calculate $F_{N}$, and by 
averaging over $10^{4}$ runs we arrive at a reliable value for
$\left\langle F_{N}\right\rangle $, Eq.\,(\ref{eq:Fquerex}). The
results obtained with random references are depicted in Fig.\,\ref{fig:Fidelity-maximization}.

As one can see, the average fidelity $\left\langle F_{N}\right\rangle $
grows with the length $N$ of the measurement sequence. Hence in principle
an estimation based on a symmetry comparison with reference states
works. However, the average fidelities are quite low. Therefore, we
ask whether this quantum estimation method allows for adaption, which
means choosing appropriate references $\ket{r_{\nu}}$ from one measurement
step to the next.

Our goal is to maximize the fidelity of the finally estimated state.
We can directly use this as a quality measure for finding adapted
reference states. An optimal way to do this would be to calculate
the fidelity of the last estimated state and maximize it with respect
to the reference state we want to adapt. In order to keep the computational
effort reasonable, we restrict ourselves to the maximization of the fidelity
$F_{\nu+1}$ after the next measurement.

To explain this in detail, let us assume that we have already performed
$\nu$ measurements. Then we can calculate an \emph{expected} fidelity
for the next measurement step $\nu+1$ \emph{before} we perform it.
Here {}``expected'' means to average over the possible measurement
outcomes. Such a quantity is a function of the next reference state
$\left|r_{\nu+1}\right\rangle $. To get the best possible fidelity
after the next measurement we have to choose a reference that maximizes
it. 

To quantify this expected fidelity, we have to determine the states
$\left|\psi_{\nu+1,a}^{\mathrm{est}}\right\rangle $ and $\left|\psi_{\nu+1,s}^{\mathrm{est}}\right\rangle $
that we would estimate if we got the outcomes $a$ or $s$, respectively.
Note that these states themselves depend on $\left|r_{\nu+1}\right\rangle $.
Then we average over the corresponding fidelities $\left|\braket{\psi_{\nu+1,a}^{\mathrm{est}}}{\psi}\right|^{2}$
and $\left|\braket{\psi_{\nu+1,s}^{\mathrm{est}}}{\psi}\right|^{2}$.
Moreover, since we do not know $\left|\psi\right\rangle $, we have
to average over it too, and hence arrive at the condition \begin{align}
\int\mathrm{d}\Omega_{\psi}\Big\{p_{a}\left(\left|r_{\nu+1}\right\rangle ,\left|\psi\right\rangle \right)\cdot\left|\braket{\psi_{\nu+1,a}^{\mathrm{est}}}{\psi}\right|^{2}+\nonumber \\
+p_{s}\left(\left|r_{\nu+1}\right\rangle ,\left|\psi\right\rangle \right)\cdot\left|\braket{\psi_{\nu+1,s}^{\mathrm{est}}}{\psi}\right|^{2}\Big\} & \longrightarrow\mathrm{max}.\label{eq:iFM2}\end{align}
for the next reference state $\left|r_{\nu+1}\right\rangle $.

Note that we have neglected the information on the distribution of
$\ket{\psi}$, we had already obtained in the preceding steps, namely
the likelihood $L_{\nu}$, Eq.\,(\ref{eq:Like}). In other words,
the $\nu${\LARGE{} }measurement outcomes{\LARGE{} }and their corresponding
reference states, i.\,e. the sequence $\mathbb{S}_{\nu}$, enter
implicitly via the adaption of the reference states but not explicitly
in a weight factor for $\dd\Omega_{\psi}$. We gain the advantage
that all expressions to be integrated are known functions of the involved
quantum states. Hence we can easily perform the integration and arrive
at an analytical expression depending on $\left|r_{\nu+1}\right\rangle $.
Nevertheless, here might be room for further improvements of a quantum
adaption. To assess this adaption strategy, we again simulate the
corresponding estimation scheme, and compare it to the random reference
strategy. As one can see from Fig.\,\ref{fig:Fidelity-maximization},
application of the adaptive scheme gives significantly better fidelities
than a random selection of reference states. 

To further elucidate the convergence of the adaptive scheme, the evolution
of estimated states on the Bloch sphere is depicted in Fig.\,\ref{fig:EstBK}.
As expected for an adaption, the estimated states accumulate near
the unknown state. The simulation shows, that in doing so, they follow, 
within the numerical errors, an exponential distribution.

Another interesting question is the distribution of reference states.
If they would prefer a certain alignment relative to the unknown state,
e.\,g. parallel to them, an experimentalist would not have to use
the quite involved adaption procedure described above, but could simply
aim for this alignment. As can be seen from Fig.\,\ref{fig:RefBK},
this is not the case. The reference states have a clear tendency to
be on the same hemisphere as the unknown states, but they show a rather
broad distribution and do not become completely parallel.

\section{Conclusions}

Symmetry measurements have been discussed in the context of 
quantum state estimation. This measurement concept, based on a comparison, 
was inspired by those used in classical physics. 
It was shown that such a basic measurement scheme leads to a converging 
estimation of pure qubits.
Furthermore, a corresponding adaptive scheme has been formulated 
and numerically simulated.

However, the achievable fidelities do not reach the
fidelity maximum $\left\langle F_{N}^{\mathrm{opt}}\right\rangle =\frac{N+1}{N+2}$
for collective spin measurements \citep{MP95,DBE98}. This is due
to the restriction on simple symmetry considerations. In principle
it would be possible to achieve higher fidelities by further discriminating
between basis states of the symmetric subspace. However, by using
such an extension of our scheme, we would sacrifice the concept of
symmetry considerations. On the other hand it would not surpass the
adaptive schemes reported in \citep{FKF00} 
for standard projection measurements, that already are quite
close to $\left\langle F_{N}^{\mathrm{opt}}\right\rangle $.
Therefore, the presented measurement concept is not challenging 
the mentioned quantum estimation methods in any practical application.
Our aim rather was to demonstrate explicitly the simple principle of a 
symmetry comparison in quantum estimation problems.

\vspace*{1em}
\section*{Acknowledgments}

We thank Igor Jex and Ferdinand Gleisberg for various discussions.
This work was supported by the Marie Curie Research Training Network
EMALI and by a grant from the state of Baden-W\"urttemberg (LGFG).


\begin{thebibliography}{10}
\bibitem{FootMixed}Note that this is a restricted optimization. 
The reason for allowing only pure states, instead of mixed states, 
is that we want to use the a priori knowledge that the unknown state is pure.
\end{thebibliography}
\end{document}